\documentclass[
 amsmath,amssymb,twocolumn,
 aps,
prd,
]{revtex4-1}

\usepackage{graphicx}
\usepackage{dcolumn}
\usepackage{bm}

\usepackage{tikz}
\usepackage{tikz-3dplot}
\usepackage{pgfplots}
\usepackage{wrapfig}
\usetikzlibrary{patterns}
\usetikzlibrary{decorations.markings}

\begin{document}

\title{IR/UV corrections to the Casimir force}

\author{Michael~Maziashvili}
 \email{maziashvili@iliauni.edu.ge}
\affiliation{School of Natural Sciences and Engineering, Ilia State University,\\ 3/5 Cholokashvili Ave., Tbilisi 0162, Georgia}


\begin{abstract}

Hard momentum cutoff is used for estimating IR/UV corrections to the Casimir force. In contrast to the power-law corrections arising from the IR cutoff, one will find the UV cutoff-dependent corrections to be exponentially suppressed. As a consequence of this fact, there is no chance to detect the corrections due to UV cutoff arising for instance from the "minimum-length" scenarios even if fundamental quantum-gravity scale is taken around $\sim$ TeV (as is the case, for example, in various models with extra dimensions).

\end{abstract}



\maketitle

\section{Introduction}

As it was originally observed by Casimir, in presence of the parallel conducting plates, the vacuum energy of a free electromagnetic field depends on the separation of plates and gives rise to the attractive force between them \cite{Casimir:1948dh}. For the energy is a quadratic quantity in fields - one easily concludes that the effect is brought about by the quantum fluctuation of the field, which in the standard way is defined as

\begin{eqnarray}
	\delta A(x) = \sqrt{ \langle 0|A^2(x)  |0\rangle - \big(\langle 0|A(x) |0\rangle\big)^2} \,=\, \sqrt{ \langle 0|A^2(x)  |0\rangle} ~, \nonumber 
\end{eqnarray} because the vacuum average of the field operator is zero. But the vacuum expectation value of the square of field operator is divergent and thus not enables one to simply estimate the magnitude of field fluctuations. Casimir's approach for calculating the force between two conducting plates suggests a neat way for eliminating divergences. Namely, the pressure is calculated on both sides of the plate and the net force appears to be finite \cite{1985PhyA..131..228G, 1986AnPhy.168...79G, Milonni:1988zz}. As the vacuum energy is represented by the sum over the Fourier modes, it is natural to ask - how does the force change if we impose certain IR and UV cutoffs? That is the question we want to address throughout this paper.

Similar question has been addressed in a few papers \cite{Ford:1988gt, 1990JPhA...23.2401H, 1993PhRvA..48.2962F}. In \cite{Ford:1988gt} a specific distribution function was suggested for the vacuum energy distribution, as an attempt to give a real physical meaning to the cut-off function occurring in Casimir's approach. Let us note that the use of the Abel-Plana formula for carrying out the summation of vacuum energy suggests also some distribution function which is independent of the Casimir's cut-off function \cite{Mamaev:1979um, Mamaev:1979ks, Mamaev:1979zw, Mamaev:1980jn, Mamaev:1981bq}. Yet another idea suggested by the authors of \cite{1990JPhA...23.2401H} is to use the stress-energy tensor regularized by the point-splitting method, in which the spatial coordinates are evaluated at the plate. Such stress-energy tensor depends only upon the time difference $T_{\alpha\beta}(t- t')$. It is the limit $t\to t'$ of this quantity that is related to the Casimir force. The frequency spectrum of which is defined by using the Fourier transform $\int\mathrm{d}\tau\exp(-i\omega \tau) T_{\alpha\beta}(\tau)$. None of these approaches give a direct clue to the question posed above. 

In the present paper we present more straightforward treatment of the IR/UV dependence of the Casimir force. The paper is organized as follows. In section \ref{Casimir force} the expression of Casimir force is put in the form convenient for our discussion and the corrections due to IR cutoff are estimated. In section \ref{Power-law dependence on the UV cutoff} we give a short discussion about the power-law corrections that may come from UV cutoff. Section \ref{Hard momentum cutoff} is devoted to the UV corrections coming from the hard UV cutoff, which is followed by short summary and conclusions in Section \ref{Concluding remarks}.

\section{Casimir force}
\label{Casimir force}

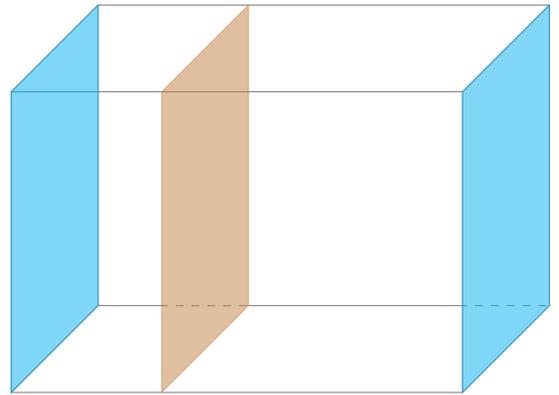
\begin{figure}[h]
\centering
\begin{tikzpicture}


\draw[gray]  (-2.5, 0, 0)  -- (-1.68, 0, 0) ;

\draw[gray, dashed]  (-1.68, 0, 0)  -- (-0.5, 0, 0) ;

\draw[gray]  (-0.5, 0, 0)  -- (2.3, 0, 0) ;

\draw[gray, dashed]  (2.3, 0, 0)  -- (3.5, 0, 0) ;

\draw[gray]
        (-2.5, 4, 0)
        -- (-0.5, 4, 0) -- (3.5, 4, 0);

\draw[gray] (-2.5, 4, 0) -- (-2.5, 0, 0) -- (-2.5, 0, 3) -- (-2.5, 4, 3) -- (-2.5, 4, 0);        
        
\draw[gray] (3.5, 4, 0) -- (3.5, 0, 0) -- (3.5, 0, 3) -- (3.5, 4, 3) -- (3.5, 4, 0);                

\draw[gray]
        (-2.5, 4, 3)
        -- (-0.5, 4, 3) -- (3.5, 4, 3);  
        
\draw[gray]
        (-2.5, 0, 3)
        -- (-0.5, 0, 3) -- (3.5, 0, 3);        
 
 
 
  


  \filldraw[cyan, opacity = .5]
       (-2.5, 0, 0)
        -- (-2.5, 4, 0)
        -- (-2.5, 4, 3)
        -- (-2.5, 0, 3)
        -- cycle;
        
        \filldraw[brown, opacity = .5]
       (-0.5, 0, 0)
        -- (-0.5, 4, 0)
        -- (-0.5, 4, 3)
        -- (-0.5, 0, 3)
        -- cycle;

        \filldraw[cyan, opacity = .5]
       (3.5, 0, 0)
        -- (3.5, 4, 0)
        -- (3.5, 4, 3)
        -- (3.5, 0, 3)
        -- cycle;

\end{tikzpicture}
\caption{Three mirrors in a box for calculating the Casimir pressure exerted upon the middle plate.} \label{Kräfte}
\end{figure}

Following the Casimir's original idea \citep{Casimir:1948dh}, for estimating the vacuum force between two mirrors placed parallel to each other at $x=0$ and $x=d$, respectively, it is expedient to introduce the third mirror at $x=l$, where it is understood that $l\gg d$, see Fig.\ref{Kräfte}.

The vacuum (otherwise named as a zero-point) energy in the region on the left-hand side of the middle mirror (we shall call it region I) is given by the sum (note that the indices $j_2, j_3$ take on all integer values)

\begin{eqnarray}
\mathcal{E}_I \,=\, \frac{1}{2}\sum_{ j_2, j_3} \sqrt{ \left(\frac{2\pi j_2}{a}\right)^2 + \left(\frac{2\pi j_3}{a}\right)^2}  \,+  \nonumber \\ \sum_{j_1=1}^\infty \sum_{ j_2, j_3} \sqrt{\left(\frac{\pi j_1}{d}\right)^2 + \left(\frac{2\pi j_2}{a}\right)^2 + \left(\frac{2\pi j_3}{a}\right)^2} ~. \end{eqnarray} Here $a$ stands for the linear size of the plate and we have tacitly assumed natural units: $c=\hbar =1$. By including the cutoff/distribution function for vacuum fluctuations and replacing summation over $j_2, j_3$ indices by the integration for $a$ is assumed to be large enough, this expression takes the form \citep{Casimir:1948dh}

\begin{eqnarray}\label{Gesamtenergie} \mathcal{E}_I  \, \approx \,   \frac{a^2}{8\pi^2} \iint \mathrm{d}k_y\mathrm{d}k_z \, f(k/\Lambda)\sqrt{k_y^2+k_z^2} \,+\, \nonumber \\ \frac{a^2}{4\pi^2} \sum_{j=1}^\infty\iint \mathrm{d}k_y\mathrm{d}k_z \, f(k/\Lambda)\sqrt{\left(\frac{\pi j}{d}\right)^2+k_y^2+k_z^2} \nonumber \\ = \, \frac{a^2}{8\pi^2} \iint \mathrm{d}k_y\mathrm{d}k_z \, f(k/\Lambda)\sqrt{k_y^2+k_z^2} \,+\, \nonumber \\ \frac{a^2}{4\pi} \sum_{j=1}^\infty\int\limits_{\pi^2j^2/d^2}^\infty \mathrm{d}\xi \, f\left(\sqrt{\xi}/\Lambda\right) \sqrt{\xi}  ~,  
\end{eqnarray} where $k^2 = \xi = (\pi j/d)^2+k_y^2+k_z^2$. The distribution function, $f(k/\Lambda)$, is introduced to regularize the energy expression at the intermediate stage and the limit $\Lambda \to \infty$ is taken at the end of calculation.

It is compelling to put the problem from the very outset in terms of pressure rather than energy \cite{1985PhyA..131..228G, 1986AnPhy.168...79G, Milonni:1988zz}. For the sake of simplicity, one can use the thermodynamic argument for its calculation. Moving a little bit the middle mirror along the $x$ axis, the variation of energy in region I will equal to the work done by the force acting on the mirror from that region: $\delta\mathcal{E} = -p \delta V$, where the change of volume is: $\delta V = a^2 \delta d$. Thus, the corresponding pressure, $p_I = -a^{-2}\partial\mathcal{E}_I/\partial d$, can be written as a sum over the Fourier modes          
 
\begin{eqnarray}
 p_I \,=\, -\, \frac{1}{a^2} \, \frac{\partial\mathcal{E}_I(d)}{\partial d}  \,=\, -\, \frac{\pi^2}{2d^4}  \sum_{j=0}^\infty  j^3 f\left(\frac{\pi j}{d\Lambda} \right) ~. \nonumber 
\end{eqnarray} Negative sign of the pressure indicates that the force acting on unit area from region $I$ is directed inwards. Analogously, the pressure in region II (the region on the right-hand side of the middle mirror) is given by 
 
 \begin{eqnarray}
p_{II} \,=\,  -\, \frac{\pi^2}{2(l-d)^4}  \sum_{j=0}^\infty  j^3 f\left(\frac{\pi j}{(l-d)\Lambda} \right) \, \approx  \nonumber \\ -\,  \frac{\pi^2}{2d^4}  \sum_{j=0}^\infty \frac{d}{l} \left(\frac{d\, j}{l}\right)^3f\left(\frac{\pi dj}{d \Lambda l} \right) \, \approx \nonumber \\ -\,  \frac{\pi^2}{2d^4}\int_{0}^\infty \mathrm{d}j \, j^3 f\left(\frac{\pi j}{d\Lambda} \right) ~, \nonumber 
 \end{eqnarray} (here we have used the fact that $l$ can be taken arbitrarily large). Hence, one finds 
  
\begin{eqnarray}\label{tsnevataskhvaoba}
p_I - p_{II} \,=\, -\, \frac{\pi^2}{2d^4} \left( \sum_{j=0}^\infty \,-\, \int_{0}^\infty \mathrm{d}j \right) j^3 f\left(\frac{\pi j}{d\Lambda} \right) ~. 
\end{eqnarray} This expression can be estimated by means of the Euler-Maclaurin formula,

\begin{eqnarray}\label{Euler-Maclaurin}
\int\limits_0^\infty \mathrm{d}j \, \mathcal{G}(j) \,-\, \frac{\mathcal{G}(0)}{2} \,-\,\sum_{j=1}^\infty \mathcal{G}(j) \,+ \, \frac{\mathcal{G}(\infty)}{2} \, = \,    \nonumber \\  -\, \left.\frac{\mathcal{G}^{(1)}(j)}{12}\right|_0^\infty \,+\, \left.\frac{\mathcal{G}^{(3)}(j)}{30\cdot 4!}\right|_0^\infty \,-\,   \left.\frac{\mathcal{G}^{(5)}(j)}{42\cdot 6!}\right|_0^\infty \nonumber \\ \,+\, \left.\frac{\mathcal{G}^{(7)}(j)}{30\cdot 8!}\right|_0^\infty \,-\, \left.\frac{5\mathcal{G}^{(9)}(j)}{66\cdot 10!}\right|_0^\infty \,+\, \cdots ~, ~~~~~~
\end{eqnarray} which enables one to state that the result is independent of a particular form of distribution function \cite{Casimir:1948dh}. However, instead of calculating the net pressure, $p_I - p_{II}$, it gives more physical insight to estimate each term separately. For this purpose let us use the cutoff-function $f(k_j/\Lambda) = \exp(-k_j/\Lambda)$ first considered in \cite{Fierz:1960zq}. Then the sum and integral entering the Eq.\eqref{tsnevataskhvaoba} can be estimated as

\begin{eqnarray}
\sum_{j=0}^\infty j^3 \exp\left(-\, \frac{\pi j}{d\Lambda} \right) \,=\, \frac{6(d\Lambda)^4}{\pi^4} \,+\, \frac{1}{120} \,-\, \frac{\pi^2}{504 (d\Lambda)^2} \,+\, \nonumber \\  \frac{\pi^4}{5760(d\Lambda)^4} \,+\, O\left(\frac{1}{(d\Lambda)^6}\right) ~, \nonumber 
\end{eqnarray} and

\begin{eqnarray}
\int\limits_0^\infty \, j^3 \exp\left(-\, \frac{\pi j}{d\Lambda} \right) \,=\, \frac{6(d\Lambda)^4}{\pi^4} ~. \nonumber 
\end{eqnarray} So that the divergences cancel out and the net pressure takes the form

\begin{eqnarray}\label{uvshestsorebatsnevaze}
p_I - p_{II} \,=\, -\, \frac{\pi^2}{240d^4} \,+\, \frac{\pi^4}{1008 d^6\Lambda^2} \,+\, O\left(\frac{1}{d^8\Lambda^4}\right) ~.
\end{eqnarray} Roughly speaking, as the cutoff-function causes the exponential cutoff of the modes $k\gtrsim \Lambda$, this result can be interpreted as the Casimir force provided by the modes $k\lesssim \Lambda$ (under assumption that $(d\Lambda)^2\gg 1$). But, as it is plain to see, this result depends crucially on the choice of cutoff function, see section \ref{Power-law dependence on the UV cutoff}.

The IR cutoff dependence is unambiguous. Namely, eliminating the modes $k<k_c$, one obtains

\begin{widetext}
\begin{eqnarray}
\sum_{j=dk_c/\pi}^\infty j^3 \exp\left(-\, \frac{\pi j}{d\Lambda} \right) \,=\, \frac{(dk_c/\pi)^3\mathrm{e}^{-k_c/\Lambda}}{1 - \mathrm{e}^{-\pi/d\Lambda}} \,+\, \frac{\Big(3(dk_c/\pi)^2 \,+\, 3dk_c/\pi \,+\, 1\Big)\mathrm{e}^{-k_c/\Lambda - \pi/d\Lambda}}{(1 - \mathrm{e}^{-\pi/d\Lambda})^2} \,+\, \nonumber \\ \,+\, \frac{\Big( 6dk_c/\pi + 6\Big)\mathrm{e}^{-k_c/\Lambda- 2\pi/d\Lambda}}{\left(1 - \mathrm{e}^{-\pi/d\Lambda}\right)^3} \,+\, \frac{6\mathrm{e}^{-k_c/\Lambda- 3\pi/d\Lambda}}{\left(1 - \mathrm{e}^{-\pi/d\Lambda}\right)^4} \,=\, \frac{6d^4\Lambda^4}{\pi^4} \,+\, \nonumber \\ \frac{1}{120}\left( 1 \,-\, 30\left(\frac{dk_c}{\pi}\right)^2 \,+\, 60\left(\frac{dk_c}{\pi}\right)^3 \,-\,  30\left(\frac{dk_c}{\pi}\right)^4  \right) \,-\,  \frac{dk_c}{30\pi}\left( 1 \,-\, 10\left(\frac{dk_c}{\pi}\right)^2 \,+\, 15\left(\frac{dk_c}{\pi}\right)^3 \,-\,  6\left(\frac{dk_c}{\pi}\right)^4  \right)\frac{\pi}{d\Lambda} \,-\, \nonumber \\ \frac{\pi^2}{504 d^2\Lambda^2} \left( 1 \,-\, 21\left(\frac{dk_c}{\pi}\right)^2 \,+\, 105\left(\frac{dk_c}{\pi}\right)^4 \,-\,  126\left(\frac{dk_c}{\pi}\right)^5 \,+\, 42 \left(\frac{dk_c}{\pi}\right)^6 \right) \,+\, O\left(\frac{1}{d^3\Lambda^3}\right) ~, \nonumber \\ \int\limits_{dk_c/\pi}^\infty \mathrm{d}j \, j^3 \mathrm{e}^{-\pi j /d\Lambda} \,=\,  \frac{d^4\Lambda^4\mathrm{e}^{-k_c/\Lambda}}{\pi^4} \left(6 \,+\, 6 \, \frac{k_c}{\Lambda} \,+\, 3\left(\frac{k_c}{\Lambda}\right)^2 \,+\, \left(\frac{k_c}{\Lambda}\right)^3 \right) \,=\, ~~~~~~~~~~~~~~~~~~~~~~~~~~~~~~~ \nonumber \\ \frac{6d^4\Lambda^4}{\pi^4} \,-\, \frac{d^4k_c^4}{4\pi^4} \,+\, \frac{d^4k_c^5}{5\pi^4\Lambda} \,-\, \frac{d^4k_c^6}{12\pi^4\Lambda^2} \,+\, O\left(\frac{1}{d^3\Lambda^3}\right) ~. \nonumber 
\end{eqnarray}\end{widetext} and thereby

\begin{eqnarray}
p_I \,-\, p_{II} \,=\, - \,\frac{\pi^2}{240d^4}\left( 1 \,-\, 30\left(\frac{dk_c}{\pi}\right)^2 \,+\, 60\left(\frac{dk_c}{\pi}\right)^3  \right) \,+\,  \nonumber \\  \frac{\pi^2k_c}{60d^4\Lambda}\left( 1 \,-\, 10\left(\frac{dk_c}{\pi}\right)^2 \,+\, 15\left(\frac{dk_c}{\pi}\right)^3  \right) \,+\, O\left(\frac{1}{d^6\Lambda^2} \right) ~. \nonumber 
\end{eqnarray} Note that the same result can be obtained by the Abel-Plana formula \cite{Olver}

\begin{eqnarray}\label{Abel-Plana}
\sum_{j=j_1}^{j_2}\mathcal{G}(j) \,-\, \int\limits_{j_1}^{j_2}\mathrm{d}j \, \mathcal{G}(j) \,=\, \frac{\mathcal{G}(j_1)}{2} \,+\, \frac{\mathcal{G}(j_2)}{2} \,+\, \nonumber \\ 2 \int\limits_0^\infty \mathrm{d}y \, \frac{\Im\left[\mathcal{G}(j_2 +iy) \,-\, \mathcal{G}(j_1 +iy)\right]}{\exp(2\pi y) \,-\, 1} ~. 
\end{eqnarray} Applying this formula to the Casimir force, it is usually assumed that $j_1=0, j_2=\infty$ and $\mathcal{G}(\infty) = 0$. If we set $j_1= dk_c/\pi$, then this formula for the Casimir force will result in 

\begin{widetext}
\begin{eqnarray}\label{irchamochra}
p_I \,-\, p_2 \,=\, -\, \frac{\pi^2}{d^4} \int\limits_0^\infty \frac{\mathrm{d}y \, y^3}{\exp(2\pi y) \,-\, 1} \,+\, \frac{3k_c^2}{d^2} \int\limits_0^\infty \frac{\mathrm{d}y \, y}{\exp(2\pi y) \,-\, 1} \, - \, \frac{k_c^3}{4\pi d} \,=\, -\, \frac{\pi^2}{240 d^4} \,+\, \frac{k_c^2}{8d^2} \, - \, \frac{k_c^3}{4\pi d}  ~.    
 \end{eqnarray}\end{widetext}

\section{Power-law dependence on the UV cutoff}
\label{Power-law dependence on the UV cutoff}

One must beware of assigning too much physical significance to the cutoff dependent contribution in Eq.\eqref{uvshestsorebatsnevaze}, as it essentially depends on the form of a cutoff function. For instance, taking $\exp\left(-\left(\pi j /d\Lambda\right)^4\right)$ instead of $\exp(-\pi j /d\Lambda)$ as a cutoff function - one finds by using the Euler-Maclaurin formula (see Eq.\eqref{Euler-Maclaurin})

\begin{eqnarray}
p_I - p_{II} \,=\, -\, \frac{\pi^2}{240d^4} \,+\, \frac{0.0021 \pi^6}{d^8\Lambda^4} \,+ \cdots ~. \nonumber
\end{eqnarray} In this expression, the leading-order cutoff-dependent term is essentially next-to-leading order term in Eq. \eqref{uvshestsorebatsnevaze}. Considering different cutoff functions, one can easily convince oneself that the leading cutoff dependent corrections take on different values. In view of this unambiguity, one may try to figure out some intuitive way for a qualitative estimate of the cutoff correction. For instance one can view the effect of cutoff function as a spatial averaging. The regularization of electromagnetic field energy by including a cutoff function: $f^2(\mathbf{k})$,

\begin{eqnarray}
\int\mathrm{d}^3k \, \left\{ \widetilde{\mathbf{E}}^2(\mathbf{k}) \,+\, \widetilde{\mathbf{H}}^2(\mathbf{k})  \right\}f^2(\mathbf{k}) ~, \nonumber 
\end{eqnarray} can be expressed as

 \begin{eqnarray}
 \int\mathrm{d}^3x \,  \left\{ \mathbf{D}^2(\mathbf{r})  \,+\, \mathbf{H}^2(\mathbf{r}) \right\} ~, \nonumber 
 \end{eqnarray} where

  \begin{eqnarray}
  \mathbf{D}(\mathbf{r}) \,=\, \int\mathrm{d}^3x' \, g(\mathbf{r} - \mathbf{r}')\mathbf{E}(\mathbf{r}') ~, \nonumber \\  \mathbf{B}(\mathbf{r}) \,=\, \int\mathrm{d}^3x' \, g(\mathbf{r} - \mathbf{r}')\mathbf{H}(\mathbf{r}') ~. \nonumber 
  \end{eqnarray} and

  \begin{eqnarray}
  g(\mathbf{r}) \,=\, \frac{1}{(2\pi)^3} \int\mathrm{d}^3k\,  f(\mathbf{k}) \mathrm{e}^{-i\mathbf{k}\cdot\mathbf{r}} ~. \nonumber 
  \end{eqnarray} Thus, $\mathbf{D}, \mathbf{B}$ are smeared out expressions for field intensities and satisfy the equations of motion

 \begin{eqnarray*}
 &&\nabla\cdot\mathbf{D} \,=\, 4\pi \langle\rho\rangle ~, ~~~  \nabla\cdot \mathbf{B} \,=\, 0 ~, \\&& \nabla\wedge \mathbf{D} \,+\, \frac{1}{c}\, \frac{\partial \mathbf{B}}{\partial t} \,=\, 0 ~,  \\ &&
 \nabla\wedge\mathbf{B}  \,-\, \frac{1}{c} \, \frac{\partial\mathbf{D}}{\partial t} \,=\, \frac{4\pi}{c} \, \langle\mathbf{J}\rangle ~.
 \end{eqnarray*} If source terms (charge and current) were initially confined to the plane, then the averaged values would reside in the slab the width of which is determined by the size of $g$ function. Hence, the distance between mirrors gets altered by the size of $g$ function, which may naturally be related to the UV cutoff as $\alpha \Lambda^{-1}$, where $\alpha$ is a certain numerical factor of order unity. Accordingly, the Casimir force gets modified as         
 
\begin{eqnarray}
-\,\frac{\pi^2}{240 d^4  } ~ \to ~ -\,\frac{\pi^2}{240 \big(d \,\pm\, \alpha \Lambda^{-1}\big)^4  } \,=\, \nonumber \\ -\,\frac{\pi^2}{240 d^4  } \left(1 \,\mp\, \frac{4\alpha}{d \Lambda} \,+\, \frac{10\alpha^2}{\left(d \Lambda\right)^2} \,\mp\, \frac{30\alpha^3}{\left(d \Lambda\right)^3} \,+\, \cdots \right) ~. 
\end{eqnarray} Thus, one may infer that the leading order correction should behave as $(\Lambda d)^{-1}$. Instead of putting in the detailed physics of the plates, which is beyond the scope of our discussion, we will proceed by estimating the UV corrections due to hard cutoff of the momentum.

\section{Hard momentum cutoff: Exponential suppression}
\label{Hard momentum cutoff}

In general, the cutoff function attaches certain weight to each Fourier mode and the Casimir pressure becomes merely a weighted average. For discussing the problem of cutoff corrections, most natural way would be to consider a hard cutoff. Hard cutoff means that a mode is either included or not. For this purpose let us use the cutoff function of the form 

\begin{eqnarray}
\frac{1}{2} \left(	1 \,-\, \tanh\left(\frac{\pi j}{d\mu} \,-\, \frac{\Lambda}{\mu} \right) \right) ~. 
\end{eqnarray} Roughly, this function changes its value from $1$ to zero in the vicinity of $\Lambda$ over the region $\mu$. Taking $\mu \to 0$, one arrives at the step-function, but $\mu$ cannot be made arbitrarily small as in Eq.\eqref{tsnevataskhvaoba} we have sum over the discrete momentum, $\pi j /d$, in Eq.\eqref{tsnevataskhvaoba}. Thus, $\mu$ is naturally defined to be of the order of $d^{-1}$. Exploiting this cutoff function and Abel-Plana formula \eqref{Abel-Plana}, one finds

\begin{eqnarray}\label{exponentsialuri}
p_I \,-\, p_2 \,=\,  -\, \frac{\pi^2}{d^4} \int\limits_0^\infty \frac{\mathrm{d}y  }{\exp(2\pi y) \,-\, 1} \, \frac{y^3}{2} \times \nonumber \\  \left( 1 \,+\, \frac{1 \,-\, \mathrm{e}^{-4\Lambda / \mu}}{1 \,+\, 2 \mathrm{e}^{-2\Lambda /\mu} \cos\left(2\pi y /\mu d\right) \,+\, \mathrm{e}^{-4\Lambda /\mu}} \right) \,=\, \nonumber \\ -\, \frac{\pi^2}{d^4} \int\limits_0^\infty \frac{\mathrm{d}y \, y^3 }{\exp(2\pi y) \,-\, 1}  \left(1 \,-\, \mathrm{e}^{-2\Lambda / \mu}\cos\left(\frac{2\pi y}{\mu d}\right) \,-\, \right. \nonumber \\ \left. \mathrm{e}^{-4\Lambda /\mu} \,+\, 2\mathrm{e}^{-4\Lambda / \mu} \cos^2\left(\frac{2\pi y}{\mu d}\right) \,+\, \cdots  \right) ~. ~~~
\end{eqnarray} Recalling that $\mu \simeq d^{-1}$, one infers that the leading UV correction to the force is controlled by the term $\mathrm{e}^{-\Lambda d}$. Hence, the contribution of high frequency modes to the Casimir force is exponentially suppressed

\section{Concluding remarks}
\label{Concluding remarks}

\begin{figure}
\begin{tikzpicture}
\begin{axis}[
xlabel={Parameter $\alpha$},
ylabel={Casimir pressure in units of $d^{-4}$},
ymajorgrids=true,
grid style=dashed,
] 
\addplot[domain=0:1.58, 
samples=100, color=cyan, thick]{-x^3/(4*pi) +x^2/8 -pi^2/240 };
\end{axis}
\end{tikzpicture}\caption{Plot of Eq.\eqref{irchamochra} in units of $d^{-4}$.}\label{irmodifikatsia}\end{figure}
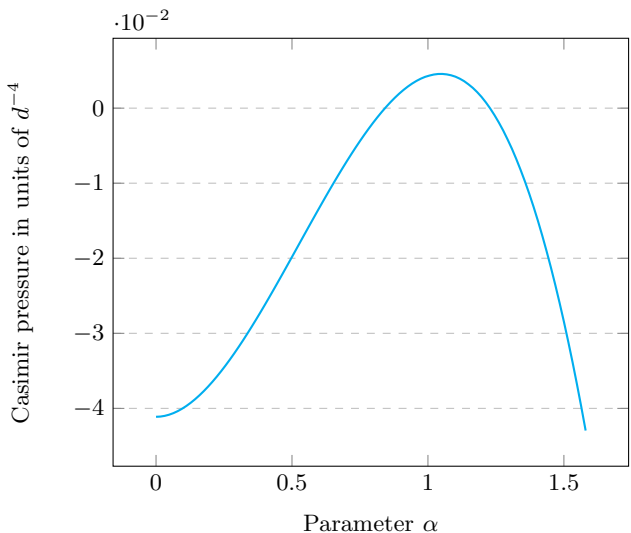

Putting the results together, one sees that Casimir force is quite sensitive to the IR cutoff and almost insensitive to the UV cutoff. Putting $k_c = \alpha /d$, from Eq.\eqref{irchamochra} it follows that the Casimir force decreases as $\alpha$ increases from $0$, in the interval $ 0.842 \lesssim \alpha \lesssim 1.228$ the force becomes repulsive, then it becomes again attractive and increases as it is depicted in Fig.\ref{irmodifikatsia}.

We can now make the connection between the above discussion and a possible corrections to the Casimir force due to minimum-length deformation of quantum theory \cite{Harbach:2005yu, Nouicer:2005dp, Frassino:2011aa}. Loosely speaking, The implementation of minimum length in quantum mechanics can be done either by modification of position and momentum operators or by the restriction of their domains. The latter possibility (see for example \cite{Sailer:2013yk, Sailer:2014hea}) is somewhat advantageous over the minimum-length deformation of Weyl-Heisenberg algebra \cite{Kempf:1994su} as in the latter case one faces unacceptably large effects in a classical limit \cite{Silagadze:2009vu, Maziashvili:2012zr}. In general, such theories imply modified dispersion relation and momentum cutoff set by the quantum gravity scale. As we mentioned, the former feature is not necessary for implementing the concept of a minimum-length into quantum theory. To work out the corrections to the Casimir force due to modified dispersion relation is straightforward. As to the corrections because of UV cutoff of the momentum, they are exponentially suppressed with respect to the Eq.\eqref{exponentsialuri} and therefore there is no chance of their detection. Namely, put in terms of a length, the quantum gravity scale is set by the Planck length $\approx 10^{-33}$cm. On the other hand, recalling that Casimir force between the parallel plates is measured for a separation distance $1\simeq \mu$m \cite{Bressi:2002fr}, the suppression factor becomes of the order of $\exp(-10^{28})$. Similarly, one concludes that the UV corrections are strongly suppressed by the factor $\mathrm{e}^{-\Lambda d}$ even for $\Lambda \simeq $ 1TeV. That is the quantum gravity scale in various extra-dimensional models \cite{Ponton:2012bi}.

It should be pointed out that in order to explore the implications of momentum cutoff properly, one has to discuss its impact on the plates as well. On the other hand, one may try to figure out a particular cutoff function involved in the Casimir force, which will have a physical meaning of probability distribution that the appropriate frequency modes will be confined within the plates.

\begin{acknowledgments}
Useful discussions with Zurab Kepuladze are kindly acknowledged. 
\end{acknowledgments}

\end{document}